\documentstyle[12pt]{article}

\def\abstract#1{\vskip 7mm 
	\begin{center}{\large Abstract}\par \bigskip
		\begin{minipage}[c]{12cm}
			\small #1
		\end{minipage}
	\end{center}
}
\def\title#1{\begin{center}{\Large\bf #1}\end{center}}
\def\author#1{\vskip 5mm \begin{center}{#1}\end{center}}
\def\address#1{\begin{center}{\it #1}\end{center}}

\newcommand{\bfr}{\begin{flushright}}
\newcommand{\efr}{\end{flushright}}

\begin{document}

\vspace*{-0cm}

\title{Soliton formulation by Moyal algebra}\author{Takao Koikawa\footnote{e-mail address: koikawa@otsuma.ac.jp}}
\vspace{1cm}
\address{
  School of Social Information Studies\\
         Otsuma Women's University\\
         Tama, Tokyo 206-0035, Japan\\
}
\vspace{2.5cm}
\abstract{ }
We formulate the soliton equations on the lattice in terms of the reduced Moyal
algebra which includes one parameter. The vanishing limit of the
parameter leads to the continuous soliton equations.

\newpage
It is well-known that the soliton equations are expressed as a
zero-curvature equation emerging as a compatibility condition of the
scattering problem equation and the time evolution equation in the inverse
scattering problem. The potentials are $sl(N,C)$ valued. Many of the 
familiar soliton equations fall into $N=2$ case. In this letter we shall 
consider a formulation of soliton equation by using the Moyal algebra\cite{Moy}
which is identical to $su(N)$ algebra for some value of
parameter in it\cite{Hop,FFZ,FZ}. In order to realize this, we 
replace the commutation relation in the zero-curvature equation by the
Moyal algebra, which introduces two new variables into the equations. We
assume that the potentials are expanded 
by powers of $e^p$ where the variable $p$ is one of the newly introduced
variables. Specification of the expansion determines the type of soliton 
equation. Substituting the potentials into the zero-curvature equation, 
we obtain various soliton equations. Taking out coefficients at each order
of power of $e^p$ in the zero-curvature equation, we
obtain soliton equations on the lattice, one of which is the well-known
Toda lattice equation. In general, the equations thus obtained include
one parameter originating from the Moyal algebra. It is interesting to
note that the 
parameter of the Moyal bracket has a physical meaning in the soliton
equation as spacing between particles on the lattice\cite{Koi1}. We then naturally
obtain the continuous correspondence of the discrete soliton equations
by taking the vanishing limit of the parameter.

We first recapitulate the notations of the Moyal algebra following
Strachan\cite{Str}. Define the star product by
\begin{equation}
f\star g = \exp \Bigg[ \kappa \epsilon^{ij}
\frac{\partial~}{\partial x^i}\frac{\partial~}{\partial\tilde x^j}
\Bigg] f({\bf x}) g({\bf \tilde x}) \vert_{{\bf x} = {\bf\tilde x}},
\label{eq:star1}
\end{equation}
and the brackets by 
\begin{equation}
\{f,g\}_{\pm}=\frac{f \star g \pm  \star f}{\kappa},
\end{equation}
where ${{\bf x}=(x^0,x^1)}$ and ${{\bf \tilde x}=(\tilde x^0,\tilde x^1)}
$.
Here the ''$-$'' bracket is the Moyal bracket.
We also introduce 
\noindent $D_x$ operator by 
\begin{equation}
D_x f  \bullet g = \frac{df}{dx}g - f\frac{dg}{dx}.
\end{equation}
The $D_x$ is the Hirota operator. So far we have not yet assumed 
any ${\bf x}$ dependence of the functions in the Moyal algebra. Here we
assume that the functions are the product of a function of $x$ and the
power of $e^p$.  It is interesting to note that the star product for
those functions are related to the Hirota operator;
\begin{equation}
[e^p f(x)] \star [e^p \tilde f(x) ] = e^{2p} \exp(\kappa D_x) f \bullet \tilde f.
\end{equation}
In more general cases with other $e^p$ dependences, we have
\begin{eqnarray}
&{}&[e^{mp} f(x)] \star [e^{np} g(x) ] \nonumber \\ 
&=& \exp\{(m+n)p\}\Bigg[\{\exp(n \kappa \frac{d}{dx})f(x)\}\{\exp(-m \kappa \frac{d}{dx})g(x)\}\Bigg]\nonumber \\
&=&\exp\{(m+n)p\}f(x+n\kappa)g(x-m\kappa),
\end{eqnarray}
where $m$ and $n$ are integers.
\noindent The brackets are explicitly obtained as
\begin{eqnarray}
&{}& \{e^{mp}f,e^{np}g\}_{\pm}\nonumber \\
&=&\frac{1}{\kappa}\exp\{(m+n)p\}\Bigg[f(x+n\kappa)g(x-m\kappa) \pm g(x+m\kappa)f(x-n\kappa)\Bigg].
\label{eq:bra}
\end{eqnarray}
We can naturally obtain the $\kappa \to \infty$ limit of the brackets;
\begin{eqnarray}
\lim_{\kappa \to 0} \{e^{mp}f,e^{np}g\}_{-}&=&2(n\frac{df}{dx}g-m\frac{dg}{dx}f)\exp\{(m+n)p\},
\label{eq:mlim}
\\
\lim_{\kappa \to 0} \kappa \{e^{mp}f,e^{np}g\}_{+}&=&2fg.
\label{eq:plim}
\end{eqnarray}

We shall next consider the soliton equation. Of the various formulations of
soliton theory, we consider the inverse scattering method where the
soliton equation is obtained as a compatibility condition of the
scattering problem equation and time evolution equation. The soliton
equation has a geometrical meaning as the zero-curvature equation:
\begin{equation}
\frac{\partial A_{\nu}}{
\partial x^{\mu}}-\frac{\partial A_{\mu}}{\partial x^{\nu}}+[A_{\mu},A_{\nu}]=0,
\end{equation}
where the potentials $A_{\mu}=A_{\mu}(x_0,x_1)$($\mu=0,1$) are $sl(N,C)$
valued fields 
and include a spectral parameter in the soliton theory. The expansion of
potentials in terms of the spectral parameter implies the equations at
each order of the parameter which leads to various soliton equations.

We shall discuss the Moyal version of the above equation. Our discussion 
is based on the fact that the Moyal algebra and the $su(N)$ algebra are
identical by some assumption on the parameter in the Moyal algebra\cite{Hop,FFZ,FZ}. This
means that we might replace the commutation relation by the 
Moyal algebra. The equation reads
\begin{equation} 
\frac{\partial A_{\nu}}{\partial x^{\mu}}-\frac{\partial A_{\mu}}{\partial x^{\nu}}+\{A_{\mu},A_{\nu}\}_{-}=0.
\label{eq:Moyal_zero} 
\end{equation}
In general, the replacement of the commutation relation by the Moyal
bracket introduces two new variables, say, $x$
and $p$ and so transfers the original equation to the 2 dimensional
higher equation in compensation of the loss of matrix. Here we assume that
the functions constituting the above equation (\ref{eq:Moyal_zero}) are
written as the product of function of induced variable $x$ and a function 
of another induced variable $p$ as   
\begin{eqnarray} 
A_0(x_0,x_1;x,p)&=&\sum_{k} e^{kp}a_k(x_0,x_1;x),\\
A_1(x_0,x_1;x,p)&=&\sum_{l} e^{lp}b_l(x_0,x_1;x).
\end{eqnarray}
Here $k$ and $l$ run over positive and/or non-positive integers of
which the choice determines the soliton equation. The
specification of the range of $k$ and $l$ together with the coefficients 
leads to various soliton equations. The expansions are
to be compared with the inverse scattering method where the potentials 
are expanded in terms of spectral parameter.  Here $e^p$ plays the role
of spectral parameter. The zero-curvature equation (\ref{eq:Moyal_zero})
implies a 
set of equations taken out at each order of power of $e^p$. Therefore,
in the transition to the Moyal version of soliton equation,  we
do not add two dimensions but one dimension to the original equation.  

In our previous paper\cite{Koi1}, we obtained the algebra composed of
brackets between operators which are the exponentiation of the
operators satisfying conformal algebra. The brackets are found to be
identical to those derived from the Moyal algebra by the reduction
assuming that functions of $x$ and $p$ are expanded by $e^0$ and $e^{\pm
p}$. The restriction of functions to those $e^p$ dependences implies the
limitation on the 
potentials that they should take value only on the Cartan
and simple algebra. In the following examples (i) and (ii) fall into
this case. But we can allow for other $e^p$ dependences.  The example
could be found in (iii).

We shall put an ansatz that $A_{\mu}=A_{\mu}(r;x,p)$ or
$A_{\mu}=A_{\mu}(t;x,p)$ where $x_0=0$ and $x_1=r$ just to identify our
examples with the well-known equations. This leads to either
\begin{equation}
\frac{d A_0}{dr}=\{A_0,A_1\}_{-},
\label{eq:radeq}
\end{equation}
or 
\begin{equation}
\frac{d A_1}{dt}=\{A_1,A_0\}_{-},
\label{eq:timeeq}
\end{equation}
respectively. 
We shall consider several examples below.

\vspace{5mm}
\noindent (i)Toda lattice equation

\vspace{5mm}
\noindent Assume that
\begin{eqnarray}
A_1(t;x,p)&=&b(t;x)+\frac{1}{2}a(t;x)(e^{p}+e^{-p}),\\
A_0(t;x,p)&=&\frac{1}{2}a(t;x)(-e^{p}+e^{-p}).
\end{eqnarray}
The  equation (\ref{eq:timeeq}) implies
\begin{eqnarray}
\frac{d}{dt}b(t;x)&=&\frac{1}{2}\{a e^p,a e^{-p}\}_{-},\label{eq:Toda1}\\
\frac{d}{dt}a(t;x)e^{\pm p}&=&\pm\{a e^{\pm p},b e^0\}_{-}.\label{eq:Toda2}
\end{eqnarray}
By using (\ref{eq:bra}), these are
\begin{eqnarray}
\frac{d}{dt}b(t;x)&=&-\frac{1}{2\kappa}\Bigg[a^2(t;x+\kappa)-a^2(t;x-\kappa)\Bigg],\\
\frac{d}{dt}a(t;x)&=&\frac{1}{\kappa}\Bigg[b(t;x+\kappa)-b(t;x-\kappa)\Bigg]a(t;x).
\end{eqnarray}
These equations lead to
\begin{equation}
\frac{d^2}{dt^2}\rho(t;x)=\frac{1}{\kappa^2}\Bigg[e^{-\rho(t;x+2\kappa)}-2e^{-\rho(t;x)}+2e^{-\rho(t;x-2\kappa)}\Bigg],
\end{equation}
where
\begin{equation}
\rho(t;x)=-2\log a(t;x).
\end{equation}
When we impose the condition $x=n \kappa,(n \in Z)$ on the above
equation, we find that the equation is to be regarded as Toda lattice
equation except at end points where the equations are different from the
equation. This shows an important aspect of the Moyal version of soliton
equation. The Toda lattice equation can be written in terms of the Cartan
matrix which reflects the algebra we choose. The Toda lattice equation
for finite size of particles corresponds to the choice of finite
dimensional algebra $A_n$. The present type of 
equation is either the Toda lattice equation with infinite number of
particles or that on the link. Mathematically these are derived assuming
$A_{\infty}$ or the Kac-Moody algebra denoted by $A_{n,1}$,
respectively. On the other hand, it is known that the $\kappa 
\to 0$ limit in the Moyal algebra leads to the Poisson bracket. In
identifying the Moyal algebra with $su(N)$ algebra one sets the relation
$\kappa \propto \frac{1}{N}$, which means that finite $\kappa$
corresponds to finite $N$ i.e.finite number of particles. 
This shows that the Moyal version of soliton equation derives the
soliton equation with periodic boundary condition of which the 
algebraic specification is $A_{n,1}$.
We can take $\kappa \to 0$ limit of Eqs.(\ref{eq:Toda1}) and
(\ref{eq:Toda2}) by using Eq.(\ref{eq:mlim});
\begin{eqnarray}
\frac{\partial}{\partial t}b(t;x)&=&-\frac{\partial a^2}{\partial x},\\
\frac{\partial }{\partial t}a(t;x)e^{\pm p}&=&2a\frac{\partial b}{\partial x}e^{\pm p}.
\end{eqnarray}
We thus obtain the continuous counterpart of the Toda lattice equation
quite naturally:
\begin{equation}
\frac{\partial ^2}{\partial t^2}\rho(t;x)=4\frac{\partial^2}{\partial x^2}e^{-\rho(t;x)}.
\end{equation}

\vspace{5mm}

\noindent (ii)Bogomolny equation\cite{Bog,WG,LS}

\vspace{5mm}

\noindent  We assume that
\begin{eqnarray}
A_0(r;x,p)&=&e^0\psi(r)+\frac{1}{2}(e^p f(r;x)+e^{-p}f^*(r;x)),\\
A_1(r;x,p)&=&\frac{1}{2}(e^{-p}f^*(r;x)-e^p f(r;x)).
\end{eqnarray}
The Eq.(\ref{eq:radeq}) implies\cite{Str}
\begin{eqnarray}
e^0\frac{d \psi}{dr}&=&\frac{1}{2}\{e^p f,e^{-p}f^*\}_{-},
\label{eq:bogo1}
\\
e^p\frac{d f}{dr}&=&-\{e^0\psi,e^p f\}_{-},\\
e^{-p}\frac{d f^*}{dr}&=&\{e^0\psi,e^{-p}f^*\}_{-}.
\label{eq:bogo3}
\end{eqnarray}
Equivalently, these are written as
\begin{eqnarray}
\frac{d \psi(r;x)}{dr}&=&-\frac{1}{2\kappa}\Bigg[|f(x+\kappa)|^2-|f(x-\kappa)|^2\Bigg],\\
\frac{d f(r;x)}{dr}&=&-\frac{1}{\kappa}\Bigg[\psi(r;x+\kappa)-\psi(r;x-\kappa)\Bigg],\\
\frac{d f^*(r;x)}{dr}&=&-\frac{1}{\kappa}\Bigg[\psi(r;x+\kappa)-\psi(r;x-\kappa)\Bigg].
\end{eqnarray}
These lead to
\begin{equation}
\frac{d^2}{dr^2}\rho(r;x)=-\frac{1}{\kappa^2}[e^{-\rho(r;x+2\kappa)}-2e^{-\rho(r;x)}+2e^{-\rho(r;x-2\kappa)}],
\end{equation}
where
\begin{equation}
\rho(r;x)=-\log |f(r;x)|^2.
\end{equation}
By taking $\kappa \to 0$ limit in the Moyal brackets, we can
rewrite the Eqs.(\ref{eq:bogo1})-(\ref{eq:bogo3}) by using the derivative
operators instead of the differences. We thus obtain the continuous limit
of the Bogomolny equation\cite{Koi1,Koi2,BF} as 
\begin{equation}
\frac{\partial^2}{\partial r^2}\rho(r;x)=-4\frac{\partial^2}{\partial x^2}e^{-\rho(r;x)},
\end{equation}
which can be interpreted as $su(\infty)$ Bogomolny equation.

\vspace{5mm}
\noindent (iii)KM equation
\vspace{5mm}

\noindent This example is the case where the expansion of potentials
include $e^{\pm 2p}$ terms besides $e^{\pm p}$ terms, which means that
these expansion does not have an 
algebra counterpart taking value only on the Cartan sub algebra and simple 
roots in contrast with the previous examples.  Assuming that
\begin{eqnarray}
A_1(t;x)&=&a(t;x)(e^{p}+e^{-p}),\\
A_0(t;x)&=&\kappa\Bigg[-\{a(t;x)e^p,a(t;x)e^p\}_{+}+\{a(t;x)e^{-p},a(t;x)e^{-p}\}_{+}\Bigg],
\end{eqnarray}
we obtain from (\ref{eq:timeeq})
\begin{equation}
\frac{d}{dt}a(t;x)e^{\pm p}= \{a e^{\mp p}, \kappa\{a e^{\pm p},a e^{\pm p}\}_{+}\}_{-},
\end{equation}
or
\begin{equation}
\frac{d}{dt}a(t;x)=-\frac{2}{\kappa}\Bigg[a^2(t;x+2\kappa)-a^2(t;x-2\kappa)\Bigg]a(t;x).
\end{equation}
This can be written as
\begin{equation}
\frac{d}{dt}u(t;x)=\frac{4}{\kappa}\Bigg[e^{-u(t;x+2\kappa)}-e^{-u(t;x-2\kappa)}\Bigg],
\end{equation}
where
\begin{equation}
u(t;x)=-2\log a(t;x).
\end{equation}
This is the equation discussed by Kac and van Maerbeke(KM)\cite{KM}.
We can also obtain the continuous limit;
\begin{equation}
\frac{\partial}{\partial t}u(t;x)=16\frac{\partial}{\partial x}e^{-u(t;x)}.
\end{equation}

In this letter we showed the formulation of soliton equations in terms of 
the Moyal algebra and exemplified it by well-known equations. The
specification how the potentials are expanded in terms of $e^p$ leads to
the various soliton equations on one-dimensional link lattice of which
the algebraic specification is $A_{n,1}$. We found that, in the present
formulation the role of
$e^p$ is the spectral parameter in the soliton 
equation in the inverse scattering method, and the parameter $\kappa$ in
the Moyal algebra has a physical meaning of spacing on the lattice.
In KM equation we have an anti-commutation bracket expression for one of
the potentials, which might be related to the super-symmetric
formulation of the Moyal version of zero-curvature equation. We shall
leave this for the following work.

\vspace{10mm}
\noindent
The author would like to express his gratitude to Prof.S.Saito for
discussion.

\end{document}